\documentclass[12pt]{article}
\usepackage{epsfig}
\begin{document}
\title{A Digital Switch and Femto-Tesla Magnetic Field Sensor Based on Fano 
Resonance 
in a Spin Field Effect Transistor}
\author{J. Wan, M. Cahay$\thanks{Corresponding author. E-mail: 
marc.cahay@uc.edu}$
\\Department of Electrical and Computer Engineering \\
University of Cincinnati, Cincinnati, Ohio 45221, USA\\
\\S. Bandyopadhyay \\
Department of Electrical and Computer Engineering \\
Virginia Commonwealth University,
Richmond, Virginia 23284, USA}
\date{}

\maketitle

\baselineskip=24pt

\begin{center}
{\bf Abstract}
\end{center}

\bigskip

We show that  a 
Spin Field Effect Transistor, realized with a semiconductor quantum wire channel 
sandwiched between {\it half-metallic} ferromagnetic contacts, 
can have Fano resonances in the transmission spectrum. These 
resonances appear because the ferromagnets are half-metallic, so that the Fermi 
level can be placed above the majority but below the minority spin band.  In 
that case, the majority spins 
will be propagating, but the minority spins will be evanescent. At low 
temperatures, the Fano 
resonances can be exploited to 
implement a digital binary switch that can be turned on or off with a very small 
gate 
voltage swing of few tens of $\mu$V, leading to extremely small dynamic power 
dissipation during switching. An array of 500,000 $\times$ 500,000  such 
transistors can detect ultrasmall changes in a magnetic field with a sensitivity 
of 1 femto-Tesla/$\sqrt{Hz}$, if each transistor is
biased near a Fano resonance.

\pagebreak

\section{Introduction}

Despite the fact that the first Spin Field Effect Transistor (SPINFET) was 
proposed almost two decades ago \cite{datta}, and numerous clones have appeared 
since then \cite{loss,  cartoixa, flatte}, no SPINFET has ever been 
experimentally demonstrated. The primary obstacle to experimental demonstration 
is the inability to achieve high spin injection efficiency at the interface 
between the source and channel, and high spin detection efficiency at the 
interface between the channel and drain. The spin injection efficiency is 
critical in determining the transistor performance. For example, it can be 
easily shown that the maximum ratio of the on-conductance to off-conductance of 
the transistor is \cite{trivedi}
\begin{equation}
{{G_{on}}\over{G_{off}}} = {{1 + \zeta_S\zeta_D}\over{1-\zeta_s \zeta_D}}
\end{equation}
where $\zeta_S$ is the spin injection efficiency at the source/channel interface 
and $\zeta_D$ is the spin detection efficiency at the drain/channel interface. 
Consequently, in order to achieve a on-off conductance ratio of 10$^5$, typical 
of modern transistors, the spin injection and detection efficiencies have to be 
as large as 99.9995\%, which is an extremely tall order. 

The maximum spin injection efficiency demonstrated to date is 90\% at low 
temperatures \cite{fiederling}. With that value, the conductance on-off ratio is 
a mere 9.5 - a far cry from 10$^5$. In other words, the conductance modulation 
is critically dependent on spin injection efficiency. Therefore, well-engineered 
SPINFETs should have excellent spin injection and detection efficiencies.

Half metallic ferromagnets \cite{groot,pickett,sanvito}, in which carriers at 
the Fermi level have only one 
spin (majority spin), are the optimum electrical spin injectors. They are often 
invoked as the most promising route to achieving nearly 100\% spin injection 
efficiency. Here, we report our investigation of Spin Field Effect Transistors 
with half metallic source and drain contacts. An example could be an InAs 
channel with Lanthanum Strontium Manganate (LSMO) contacts, which are known to 
have very high degree of spin polarization at low temperatures and approximate 
ideal half metals. We restrict ourselves to a quantum wire channel since it is 
known to produce maximum conductance modulation \cite{datta}. 

We have studied {\it ballistic} spin transport through this device by solving 
the Pauli equation, which yields the transmission amplitudes of both majority 
and minority spins in the source contact. Assumption of ballistic transport is 
realistic since mobility in quantum wire channels can be reasonably high. We 
also neglect spin relaxation in the channel since the Elliott-Yafet mechanism 
\cite{elliott} requires carrier scattering (which is assumed to be absent in 
ballsitic models) and the D'yakonov-Perel' mechanism \cite{dyakonov} is absent 
in a quantum wire channel where only one subband is occupied. Finally, we 
neglect any scattering at the interface between the ferromagnet and 
semiconductor, which can cause spin relaxation. 

With the above model, we have studied spin dependent transmission spectra of 
electrons through the SPINFET. Our study revealed the existence of Fano 
resonances in the transmission spectra since the Fermi level in the device can 
be located above the majority spin band in the source (which contributes 
propagating electrons), but below the minority spin band (which is evanescent). 
It is well known that such a situation causes Fano resonances \cite{fano}. We 
found that the Fano resonances are narrow and well-resolved at sufficiently low 
temperatures. We can therefore bias the transistor around such a resonance (by 
applying an appropriate dc gate voltage), so that a small change in the gate 
voltage will vary the electron injection energy around the Fano resonance and 
switch the transmission through the device from maximum to minimum. This is the 
basis of a digital `switch' which can be switched on and off with a very small 
gate voltage swing, resulting in extremely small dynamic power dissipation 
during switching.

We also considered the situation when the transistor is placed in a static 
magnetic field directed along the channel. As long as the transistor is biased 
near a Fano resonance, a small change in the magnetic field can change the 
transmission through the device (and therefore the channel conductance) from 
maximum to nearly zero. Thus, this device is capable of detecting minute changes 
in a magnetic field. We show that an appropriately designed system can detect 
magnetic field changes with a sensitivity of $\sim$ 1 femtoTesla/$\sqrt{Hz}$.

This paper is organized as follows. In the next Section, we present the theory. 
In Section 3, we present results and in Section 4, we present calculations of 
the sensitivity of a magnetic field sensor based on this principle. The 
conclusions are presented in Section 5.

\section{Theory}

Consider a Spin Field Effect Transistor (Spin FET) with  half-metallic 
ferromagnetic 
source and drain contacts,  as shown in Fig. 1(a). The channel is a quantum wire 
in which 
only the lowest subband is occupied by carriers. The conduction band 
discontinuity at the 
heterointerface between the channel and the gate insulator results in an 
effective 
electric field along the y-axis, which induces a Rashba spin orbit interaction 
in the 
channel. The ferromagnetic contacts induce a static magnetic field in the 
x-direction. 
This field can also be applied with external sources.

The  equilibrium energy band diagram 
along the channel (x-direction) is shown in Fig. 1(b). 
We adopt the Stoner model and assume that the majority and minority spin bands 
in the 
ferromagnets are 
 exchange split by an amount $\Delta$. The Fermi level is below the bottom of 
the 
minority spin band so that the spin polarization of carriers at the Fermi energy 
in the 
ferromagnets is 100\%
 (hence ``half metallic''). The majority spins from the ferromagnet will be {\it 
propagating}  and the minority spins will be {\it evanescent}.  For the sake of 
simplicity, we will neglect space charge effects and assume that  the conduction 
band 
edge in the channel is flat and invariant in the x-coordinate. Furthermore, the 
contact 
potentials at the contact/channel interface will be represented by two delta 
barriers of 
height $\Gamma$.

If we assume that the confinement potential in the z-direction is parabolic and 
given by 
$V(z)$ = $(1/2)m^* \omega_0^2 z^2$, then the approximate
energy dispersion relations of the lowest spin split subbands in the channel 
(derived 
using perturbation theory) are given by \cite{cahay1,cahay2,superlattice}
\begin{equation}
E_1 (k_x) = {{1}\over{2}} \hbar \omega + \Delta E_c + {{\hbar^2 k_x^2}\over{2
m^*}}
- \sqrt{ \left (\eta k_x \right )^2 + \left (
\frac{ g^* \mu_B B}{2} \right )^2 },
\label{dispersion1}
\end{equation}
\begin{equation}
E_2 (k_x) = {{1}\over{2}} \hbar \omega + \Delta E_c + {{\hbar^2 k_x^2}\over{2
m^*}}
+ \sqrt{ \left ( \eta k_x \right )^2 + \left (
\frac{ g^* \mu_B B}{2} \right )^2 },
\label{dispersion2}
\end{equation} 
where the subscripts `1' and `2' refer to the lower and upper subbands,  
$\omega = \sqrt{ \omega_0^2 + \omega_c^2 }$, $\omega_c=eB/m^*$, $g^*$ is the 
effective 
Land\'e g-factor in the channel,
 $k_x$ is the wavevector component in the x-direction, $m^*$ is the effective 
mass of 
carriers, and $\eta$ is the strength of the Rashba interaction. The quantity 
$\Delta E_c$ 
is the potential barrier between the ferromagnets and the channel, and includes 
the 
confinement energy due to confinement in the y-direction (see Fig. 1(b)).

If the axial magnetic field $B$ is below a critical field $B_c$ (= 4 $\eta 
m^*/(\hbar^2 g 
\mu_B)$), then the dispersion relation $E_1 (k_x)$ will have a camel-back 
shape with a local maximum at $k_x$ = 0 and two local minima at $k_x$ = $\pm 
(\sqrt{4 
\eta^2 (m^*)^2/\hbar^4 - (g^* \mu_B B)^2/4})/\eta$  \cite{superlattice}. Above 
$B_c$, the 
shape of $E_1 (k_x)$ is approximately parabolic with a global minimum at $k_x$ = 
0. The 
dispersion relation
$ E_2 (k_x)$, on the other hand, always has an approximately parabolic shape 
with a 
global minimum at $k_x$ = 0.

The eigenspinors in the two subbands $E_1(k_x)$ and $E_2(k_x)$ are 
\cite{superlattice} 
\begin{eqnarray}
\Psi_1(B, k_x) & = &
\left [ \begin{array}{c}
             cos(\theta_{k_x})\\
             sin(\theta_{k_x}) \\
             \end{array}   \right ]
\nonumber \\
\Psi_2(B, k_x) & = &
 \left [ \begin{array}{c}
              sin(\theta_{k_x})\\
               - cos(\theta_{k_x})\\
             \end{array}   \right ]
\end{eqnarray}
where $\theta_{k_x}$ = -$(1/2) arctan [(g^* \mu_B B)/(2 \eta k_x)]$.

The fact that the eigenspinors are wavevector ($k_x$) dependent tells us that 
neither 
subband has a fixed spin quantization 
axis (the spin quantization axis changes with the wavevector). At any arbitrary 
$k_x$, 
each of the two spin eigenspinors in the channel will be mutually orthogonal, 
but each 
will be a mixture (or superposition) of both majority and minority spins in the 
ferromagnetic contacts. Moreover, at any fixed incident energy, the eigenspinors 
in the 
two spin split subbands $E_1$ and $E_2$ will not be orthogonal, so that there 
will be 
coupling  between them. As a result, majority spins injected from the 
ferromagnet (with 
any arbitrary injection energy) will be mixed with minority spins in the 
channel. This is 
termed ``spin mixing''. Note that spin mixing occurs because of the {\it 
simultaneous} 
presence of the axial magnetic field and spin orbit interaction. If the former 
were 
absent, $\theta_{k_x}$ = 0, and the eigenspinors will be z-polarized states 
$\left [ 
\begin{array}{cc}
1 \\
0 \\
\end{array}
\right ]$ and $\left [ \begin{array}{cc}
0 \\
1 \\
\end{array}
\right ]$ which are wavevector independent, while if the latter were absent, 
$\theta_{k_x}$ = $\pi/2$ and the eigenspinors will be x-polarized states 
$(1/\sqrt{2})\left [ \begin{array}{cc}
1 \\
1 \\
\end{array}
\right ]$ and $(1/\sqrt{2})\left [ \begin{array}{cc}
1 \\
-1 \\
\end{array}
\right ]$ which are also wavevector independent. In both of the above cases, the 
eigenspinors in the two subbands will be always orthogonal (at any energy) and 
there will 
be no mixing or coupling between them. Therefore, simultaneous presence of the 
axial 
magnetic field and spin orbit interaction are 
required to induce spin mixing.

In order to study ballistic transport in the Spin FET, we used the model 
developed in 
refs. \cite{cahay1, cahay2} to calculate the (spin-dependent) quantum mechanical 
transmission amplitude $t$ of an electron through the channel (in the presence 
of ``spin 
mixing'') as function of the injection energy $E$. For these calculations, we 
have used 
the parameters listed in Table I. The transmission amplitude $t(E)$ also depends 
on the 
axial magnetic field $B$ and the exchange splitting $\Delta$ in the contacts. In 
\cite{cahay1, cahay2}, we assumed that the bottoms of both majority and minority 
spin 
bands in the 
ferromagnet are below the Fermi level, so that both spins are propagating modes. 
In the 
present case, we modified the formalism  to account for the fact that minority 
spins are 
evanescent.

From the calculated transmission amplitude as a function of energy, we calculate 
the 
linear response conductance using the finite temperature Landauer formula:
\begin{eqnarray}
G_{\uparrow}(\Delta E_c, B, \Delta) & = & (e^2/4hkT)\int_{\Delta E_c}^{\infty} 
dE 
|t_{\uparrow}(E, B, \Delta)|^2 sech^2 \left ({{E - E_F}\over{2 kT}} \right 
)\nonumber \\
G_{\downarrow}(\Delta E_c, B, \Delta) & = & (e^2/4hkT)\int_{\Delta E_c}^{\infty} 
dE 
|t_{\downarrow}(E, B, \Delta)|^2 sech^2 \left ({{E - E_F}\over{2 kT}} \right 
)\nonumber 
\\
G_{total}(\Delta E_c, B, \Delta) & = & G_{\uparrow}(\Delta E_c, B, \Delta) + 
G_{\downarrow}(\Delta E_c, B, \Delta)
\label{conductance} 
\end{eqnarray}
where $\uparrow$ and $\downarrow$ refer to majority and minority spins in the 
contact, 
respectively. The last equality is a consequence of the fact that in the 
ferromagnetic 
contacts, the two eigenspinors are orthogonal at any given energy. However, 
since the 
minority spin is evanescent, $G_{\downarrow}(\Delta E_c, B, \Delta)$ $\equiv$ 0, 
and 
$G_{total}(\Delta E_c, B, \Delta)$ = 
$G_{\uparrow}(\Delta E_c, B, \Delta)$. We emphasize that although the evanescent 
modes do 
not directly contribute to the total conductance, they nonetheless have an {\it 
indirect} 
influence because they renormalize the transmission probability of the 
propagating modes 
\cite{bagwell0, surface-science}, and therefore affect $G_{\uparrow}(\Delta E_c, 
B, 
\Delta)$ and ultimately $G_{total}(\Delta E_c, B, \Delta)$. 
 
Because one incident mode from the contact (majority spin) is propagating, while 
the 
other (minority spin) is evanescent, the transmission spectrum $|t_{\uparrow 
\downarrow}|$ versus $E$ will contain
resonance/antiresonance pairs closely spaced in energy,
a feature known as Fano resonances \cite{fano,bagwell1,bagwell2,serra}. From 
Equation 
(\ref{conductance}), it is obvious that these Fano resonances will show up in 
the plot of 
$G_{total}$ versus $\Delta E_c$, at sufficiently low temperatures, for a given 
value of 
$B$ and $\Delta$. Note that we can 
vary $\Delta E_c$ with the gate voltage; therefore, a plot of $G_{total}$ as 
function of 
gate voltage will show Fano resonances at low enough temperatures.

\section{Results}

In Fig. 2, we plot $G_{total}$ versus $\Delta E_c$ at a temperature of 0 K, for 
different 
values of the exchange splitting
energy $\Delta$ in the contacts. Since $\Delta E_c$ has an approximately linear 
dependence on the gate voltage $V_g$ (over small ranges of gate voltage 
variation), this 
plot can also be viewed as a plot of $G_{total}$ versus $V_g$. The magnetic 
field $B$ 
along the channel
is assumed to be 0.6 Tesla which is above the critical field $B_c$. Although 
$\eta$ 
should vary with $V_g$, this variation is negligible over the gate voltage range 
we 
consider, since $\eta$ has a weak dependence on $V_g$ \cite{nitta}. Therefore, 
we assume 
that $\eta$ is constant over the entire range of $\Delta E_c$ (or $V_g$) and  
has the 
value given in Table I.

In Fig. 2, we clearly see the Fano resonance-antiresonance pairs. Additionally, 
there are 
isolated resonances
where the conductance reaches the maximum value of $e^2/h$. These are due to 
Ramsauer 
resonances discussed in ref. \cite{cahay1}.  The locations of the Ramsauer 
resonances (on 
the $\Delta E_c$ axis)
are fairly insensitive to the exchange splitting energy $\Delta$, but
the locations of the Fano resonance-antiresonance pairs 
are strongly dependent on $\Delta$. 
This is due to the fact that  the decay length of the evanescent minority spin 
band in 
the contact has a strong dependence on $\Delta$. The value of $\Delta$ affects
the amount of coupling/mixing between the two propagating modes
in the semiconductor channel via the boundary
conditions at the contacts. That, in turn, affect the Fano resonances, but not 
the 
Ramsauer resonances.

The energy separation between the Fermi level and the subband bottoms  depends 
on the 
axial magnetic field $B$. Therefore, we expect the magnetic field to influence 
the Fano 
resonances, and it does. This is illustrated in 
Fig. 3 which is a plot of $G_{total}$ versus ${\Delta}E_c$ for different values 
of
the axial magnetic field $B$ for a fixed value of $\Delta$ = 6 eV. For the range 
of 
$\Delta E_c$ considered
here, there are two Ramsauer resonances.
In addition, there is a Fano resonance preceding each Ramsauer
resonance.  The Ramsauer resonances correspond to perfect transmission of the 
majority 
spin band, while each Fano resonance appears in place of a second
Ramsauer resonance that would have appeared owing to perfect transmission of the 
minority 
spin band, were it propagating instead of being evanescent.

Around a Fano resonance, a very small change in $\Delta E_c$ will switch the 
conductance 
from the maximum value of $e^2 /h$ to zero. Therefore, we can realize a {\it 
binary 
switch} by biasing the device close to a Fano resonance. This can be achieved  
by 
applying an appropriate fixed dc voltage on the gate.
Around the leftmost Fano resonance, the change in
$\Delta E_c$ ($\delta (\Delta E_c)$) required to switch the device on or off, is 
$\sim$ 
20 $\mu $ eV and is nearly independent of the strength of the magnetic
field $B$. Since $\delta (\Delta E_c)$ = $qV_g -qV_{ins}$ (where $V_g$ is the 
gate 
voltage, $q$ is the electronic charge and $V_{ins}$ is the gate voltage dropped 
across 
the gate insulator), the gate voltage swing $V_t$ required to switch the device 
from 
on-to-off is 20 $\mu$V + $V_{ins}$. For acceptable error rate, we want this 
voltage to 
exceed the thermal noise voltage on the gate capacitor, which is $U_{th}$ = 
$\sqrt{kT/C_g}$, where $C_g$ is the gate capacitance and $T$ is the operating 
temperature 
\cite{kish}. We will later show that the Fano resonances begin to wash out at 
temperatures above 0.1 K, so that $T$ $\leq$ 0.1 K. Assuming $C_g$ = 1 fF 
(including 
interconnects), $U_{th}$ $\leq$ 37 $\mu$V. Therefore, we must choose the gate 
insulator 
such that $V_{ins}$ $\geq$ 17 $\mu$V.

If the gate capacitance is 1 fF (including interconnects), then the energy 
dissipated in 
switching this transistor is (1/2)$CV_t^2$ = 7 $\times$ 10$^{-25}$ Joules 
$\approx$ 
$kTln2$ (at the temperature of 0.1 K). If switched adiabatically, the energy 
dissipation 
can be much less \cite{zhirnov}. Consequently, the binary switch that we propose 
here is 
an extremely {\it low power} switch.

Near a Ramsauer resonance (perfect transmission of the majority spin band), the 
conductance versus $\Delta E_c$ curve is fairly insensitive to the axial 
magnetic field 
$B$. However, near a Fano resonance, the conductance curve is extremely 
sensitive to $B$.
For instance, at zero temperature, when the gate is biased close to a Fano 
resonance (at 
a voltage such that  $\Delta E_c$ = 4.1922 eV), there is a 100$\% $ change in 
the 
conductance (from 
a maximum value of $e^2 /h $ to zero)  when the 
magnetic field $B$ is changed from 0.4 T to 0.372 T, a mere change
of 28 mT. Therefore, at very low temperatures, this device can operate as a 
magnetic 
field sensor. We show below that it can actually detect fields with a 
sensitivity of 
$\sim$ 1 femto-Tesla/$\sqrt{Hz}$, using standard analysis as in refs. \cite{aaa, 
physicaE}.

\section{Magnetic field sensor}

Consider a single device. The noise current has two components - thermal noise 
and shot 
noise. The thermal noise current is given by $I_n^{thermal}$ = $\sqrt{4 kT G 
\Delta f} 
\zeta$, where $G$ is the source-to-drain conductance, $\Delta f$ is the 
frequency 
bandwidth, and $\zeta$ is the noise suppression factor due to carrier 
confinement in 
quantum wires \cite{svizhenko}.  The shot noise current is 
$I_n^{shot}$ = $\sqrt{(2/3)e G V \Delta f}$ \cite{beenakker}. Therefore the 
total noise 
current in a transistor is $I_n$ = $C \sqrt{G}$, where $C$ = $\sqrt{[4kT \zeta^2 
+ (2/3) 
e V] \Delta f}$. 

When used as a magnetic field sensor, the change in the current through a single 
transistor in a magnetic field $H$ is the signal current $I_s$ and is given by 
$I_s$ = 
$SH$ where $S$ is the sensitivity. To keep the analysis tractable, we will 
assume that 
current changes linearly in a magnetic field, so that $S$ is independent of $H$. 
Therefore, the signal to noise ratio (SNR) for a single transistor is
\begin{equation}
SNR = {{SH}\over{I_n^{thermal} + I_n^{shot}}} = {{SH}\over{C\sqrt{G}}}
\end{equation}
The conductance of $N$ transistors operating in parallel is $NG$. Therefore, the 
noise 
current of $N$ transistors in parallel is $\sqrt{N} I_n$, whereas the signal 
current is 
$N I_s$. Thus, the SNR increases as $\sqrt{N}$. This allows us to detect very 
small 
magnetic fields by using a large number of transistors in parallel, each acting 
as a 
sensor.

For a change of magnetic field of 28 mTesla $\equiv$ 280 Oe, the conductance of 
a 
transistor changes by $e^2/h$. If the source-to-drain voltage is 100 mV, then 
the drain 
current changes by 39 $\mu$A. Therefore $S$ = 0.14 $\mu$A/Oe. For a field of 1 
femto-Tesla = 10$^{-11}$ Oe, the signal current $I_s$ in a single transistor is 
1.4 
$\times$ 10$^{-18}$ Amperes. The noise current is dominated by shot noise at the 
operating temperature of 0.1 K and is given by 6.4 $\times$ 10$^{-13}$ 
Amperes/$\sqrt{Hz}$. Therefore, the SNR for a single transistor is 2.2 $\times$ 
10$^{-6}$/$\sqrt{Hz}$. The SNR for 2.5 $\times$ 10$^{11}$ transistor sensors in 
parallel 
will be $\sqrt{2.5 \times 10^{11}}$ times larger and exceed unity/$\sqrt{Hz}$ (0 
db), 
which makes the signal measurable against the background of noise. Therefore 
with an 
array of 500,000 $\times$ 500,000 transistors, we should be able to detect a 
magnetic 
field of 1 femto-Tesla with a bandwidth of 1 Hz. With a transistor density of 
2.5 
$\times$ 10$^{9}$/cm$^2$, this sensor can be implemented  in a 10 cm $\times$ 10 
cm chip.
The dynamic power dissipated to detect 1 femto-Tesla (in a bandwidth of 1 Hz) 
will be 
only 35 nWatts.

In Fig. 4, we show that the Fano resonances wash out very quickly with 
increasing 
temperature. The Ramsauer resonances are more robust against temperature. The 
Fano 
resonances are essentially indiscernible at temperatures above 0.3 K, so that we 
should 
restrict all device operation to below 0.1 K. 

\section{Conclusion}

In conclusion, we have shown that a Spin FET with {\it half metallic} contacts 
has both 
Ramsauer and Fano resonances in the transmission spectrum. In ref. 
\cite{cahay1}, we 
found only the Ramsauer resonances, and not the Fano resonances, since we did 
not 
consider half metallic ferromagnetic contacts, so that both majority and 
minority spins 
in the contact were propagating modes. Here, we have considered the case where 
the 
majority spins are propagating, but the minority spins are evanescent. This 
causes the 
Fano resonances.
The locations of the Fano resonances are very sensitive to an axial magnetic 
field and 
the exchange splitting in the contacts. We can bias the transistor 
near a Fano resonance using a dc gate voltage and, at low temperatures, realize 
very low power binary switches (dissipating $\sim$ $kTln2$ energy per switching 
event), 
as well as sensitive magnetic field detectors that can detect ultrasmall fields 
with a 
sensitivity $\sim$ 1 femto-Tesla/$\sqrt{Hz}$.

\newpage

\vskip .1in
\begin{center}
{\bf Table I: Parameters of the spin interferometer}
\end{center}
\begin{center}
\begin{table}[h]
\centering
\begin{tabular}{cc} \hline\hline
Fermi Energy $E_F$ (eV) & 4.2         \\
Rashba spin-orbit coupling constant ${\alpha}_R$ ($10^{-11}$ eVcm)  &  1.    \\
Lande Factor $g^*$ &  -14.9   \\
Effective mass ${m_f}^* $ in Fe contact ($m_0$)  &   1.     \\
Effective mass ${m_s}^* $ in InAs channel ($m_0$)       & 0.023\\
Length of the channel(${\mu}m$) & 0.15\\
Strength of delta scatterer at the contact/channel interface
(ev $\AA$) & 2.0\\
\hline\hline
\end{tabular}
\end{table}
\end{center}

\newpage
\
\parindent 0cm
\vskip .1in
{\bf Figure 1}: (a) Structure of a Spin Field Effect Transistor with half 
metallic (HM) 
source and drain contacts. (b) Energy band diagram along the channel: 
Also shown as dashed lines are the resonant energy states above $\Delta E_c$. 
The barriers at the ferromagnet/semiconductor interface are modeled as simple 
one-dimensional delta-potentials. 

\vskip .1in
{\bf Figure 2}: 
Zero temperature conductance $G_{total}$ (=$G_{\uparrow}$) as a function of 
$\Delta E_c$ 
for various values
of the exchange energy $\Delta $ in the half-metallic contacts. The axial 
magnetic field 
$B$ =  0.6 Tesla. Other parameters used to obtain these curves are listed in 
Table I.

\vskip .1in
{\bf Figure 3}: Zero temperature conductance $G_{total}$ (=$G_{\uparrow}$) as a 
function 
of $\Delta E_c$ for various values of the axial magnetic field $B$. 
The exchange energy $\Delta $ in the half-metallic contacts is assumed to be 6 
eV. The 
Fano and Ramsauer resonances are indicated.

\vskip .1in
{\bf Figure 4}: (a) Temperature dependence of the conductance modulation of the 
Spin FET 
when the exchange energy $\Delta $ in the half-metallic contacts is 6 eV.
The Fano resonances are washed out much faster than the Ramsauer resonances
as the temperature rises. (b) Temperature dependence of the leftmost Fano 
resonance shown 
in higher resolution. From top to bottom,
the curves correspond to a temperature of 0, 0.05, 0.1, 0.15, 0.2, 0.25, 
and 0.3 K, respectively. 

\newpage
\
\vskip .2in
\begin{figure}[h]
\centerline{\psfig{figure=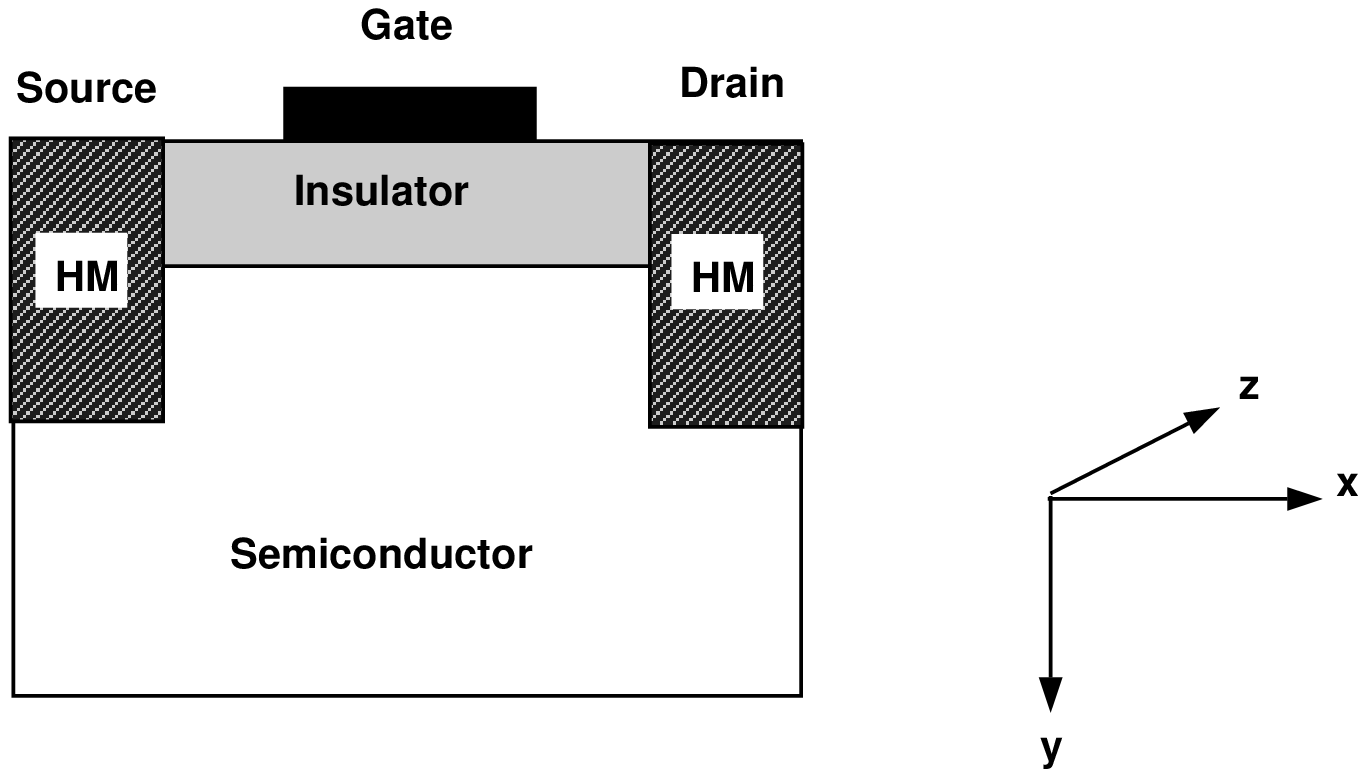,height=3in,width=5.5in}}
\end{figure}
\begin{center}
\vskip 3in
{\bf Figure 1(a)}
\end{center}

\newpage
\
\vskip .2in
\begin{figure}[h]
\centerline{\psfig{figure=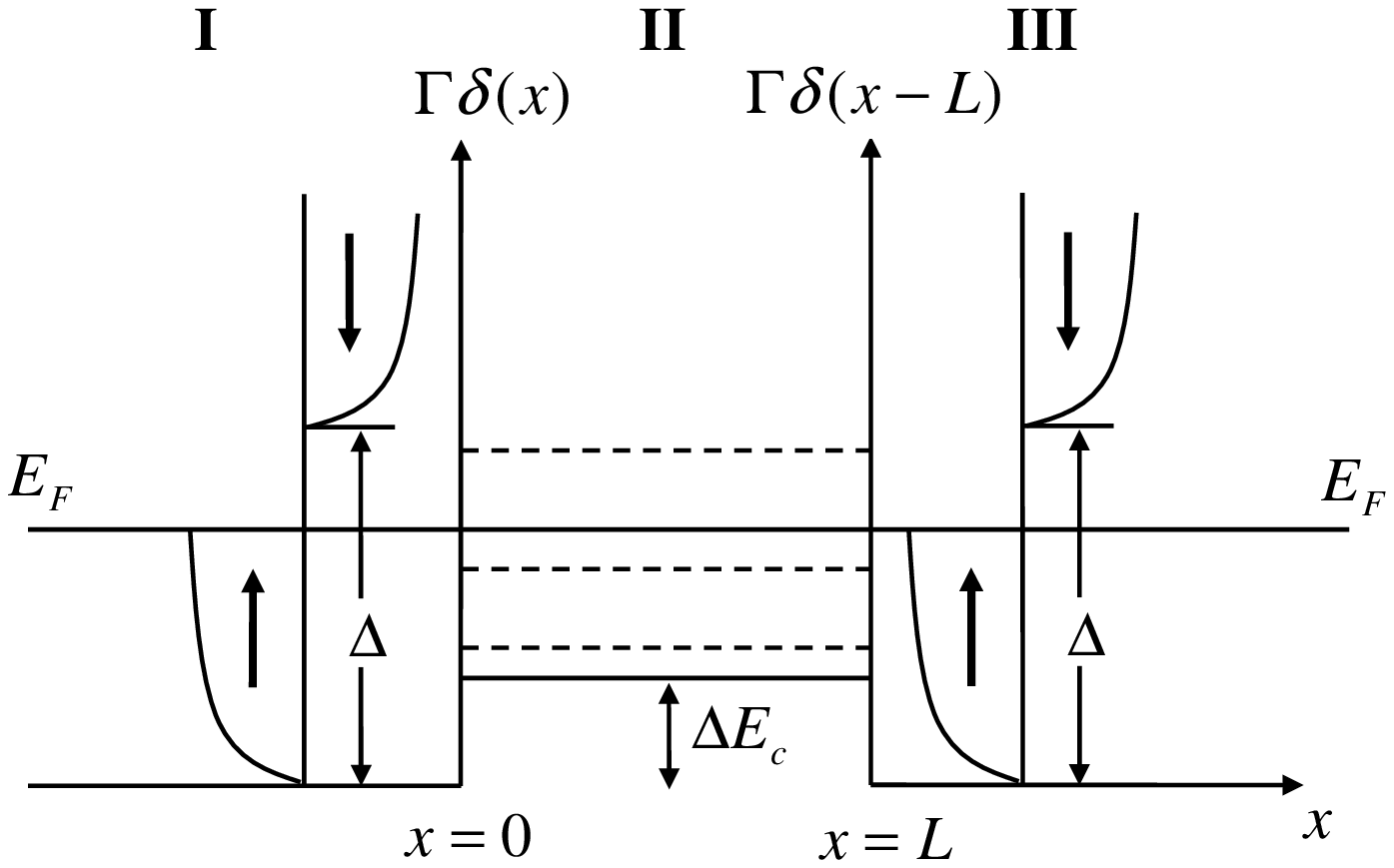,height=3in,width=5.5in}}
\end{figure}
\begin{center}
\vskip 3in
{\bf Figure 1(b)}
\end{center}

\newpage
\
\vskip 0.2in
\begin{figure}[h]
\centerline{\psfig{figure=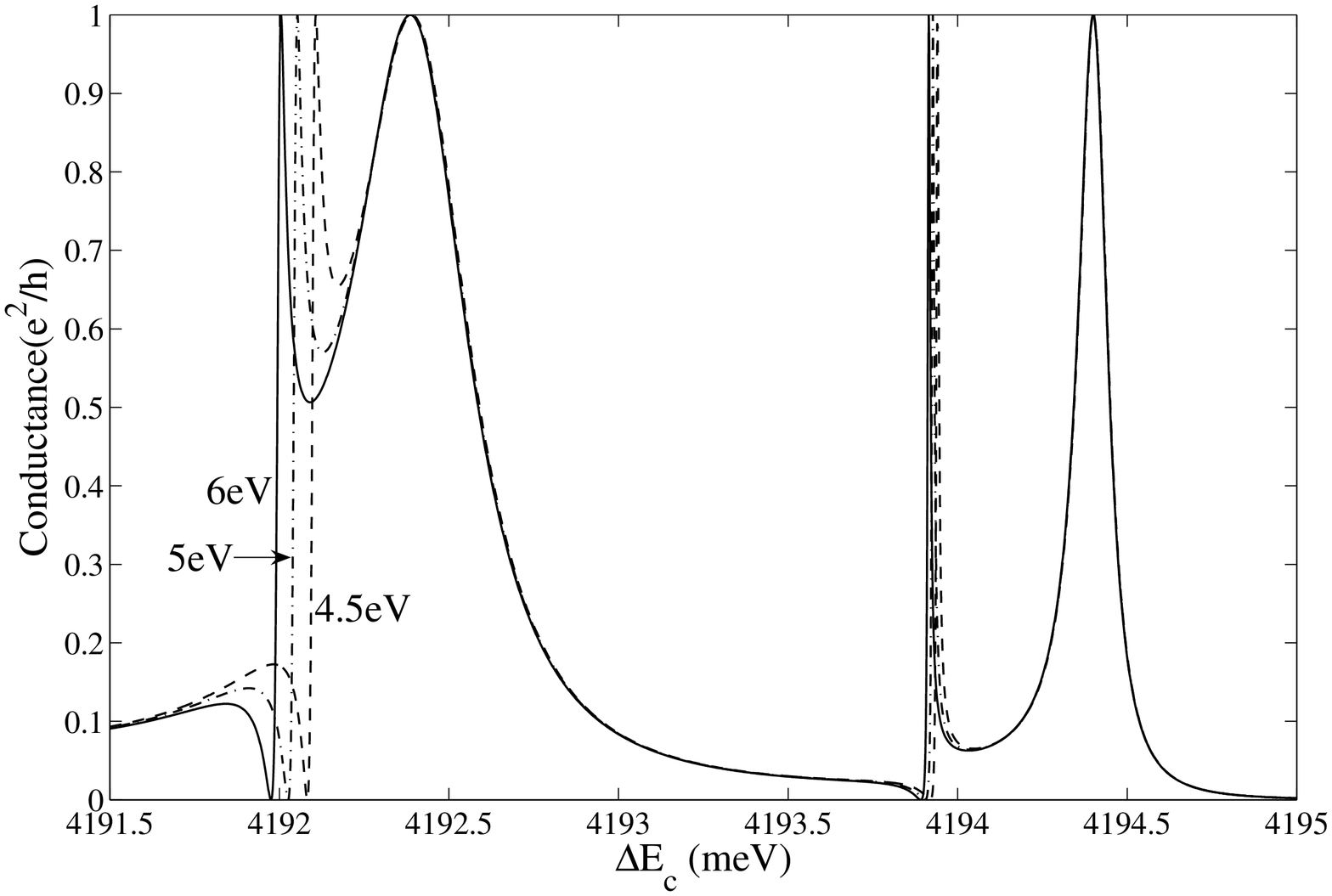,height=4.5in,width=6in,angle=-90
}}
\end{figure}
\vskip 2in
\begin{center}
{\bf Figure 2}
\end{center}

\newpage
\
\vskip 0.2in
\begin{figure}[h]
\centerline{\psfig{figure=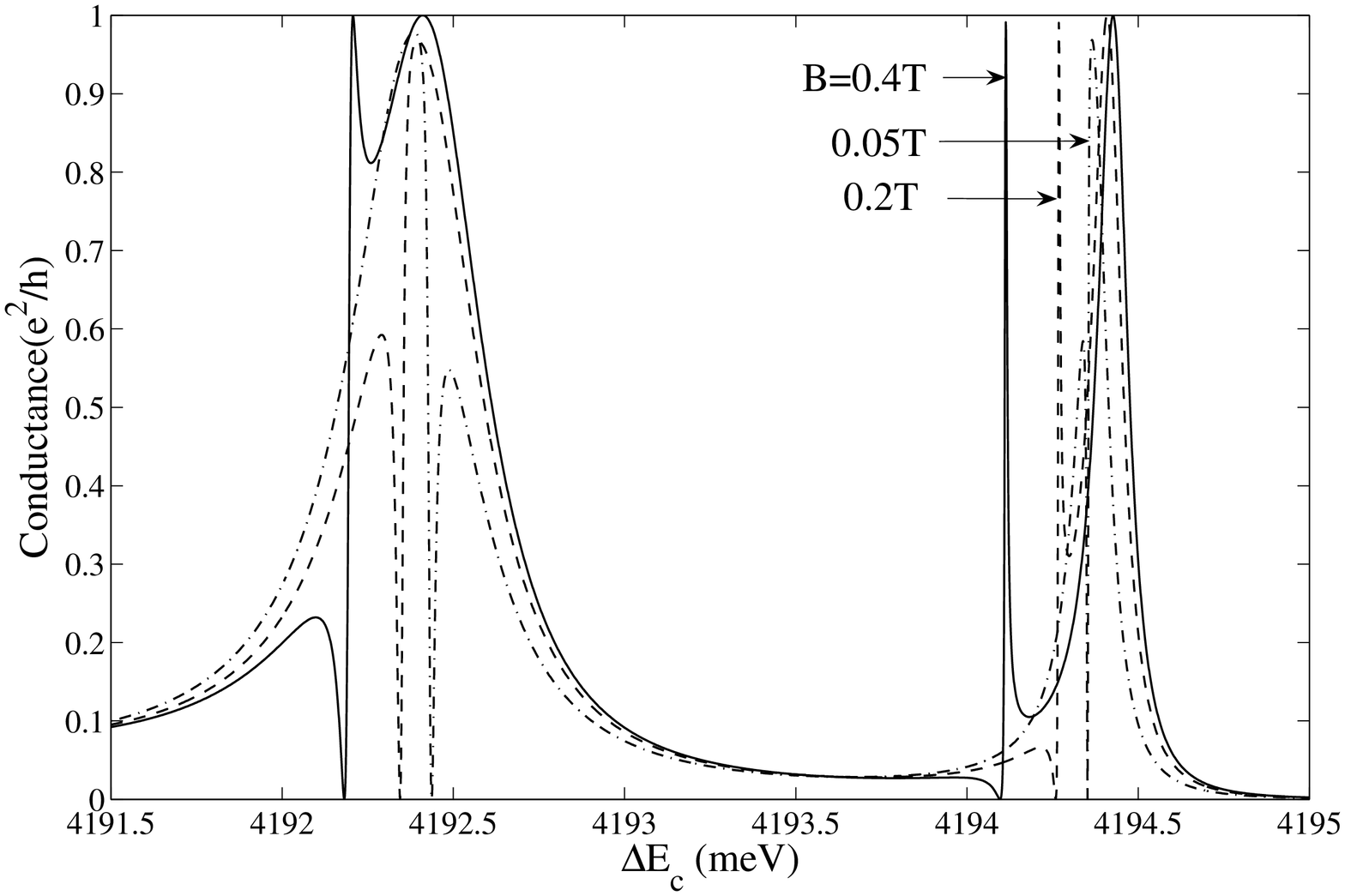,height=4.5in,width=6in,angle=-90}}
\end{figure}
\vskip 2in
\begin{center}
{\bf Figure 3}
\end{center}

\newpage
\
\vskip 0.2in
\begin{figure}[h]
\centerline{\psfig{figure=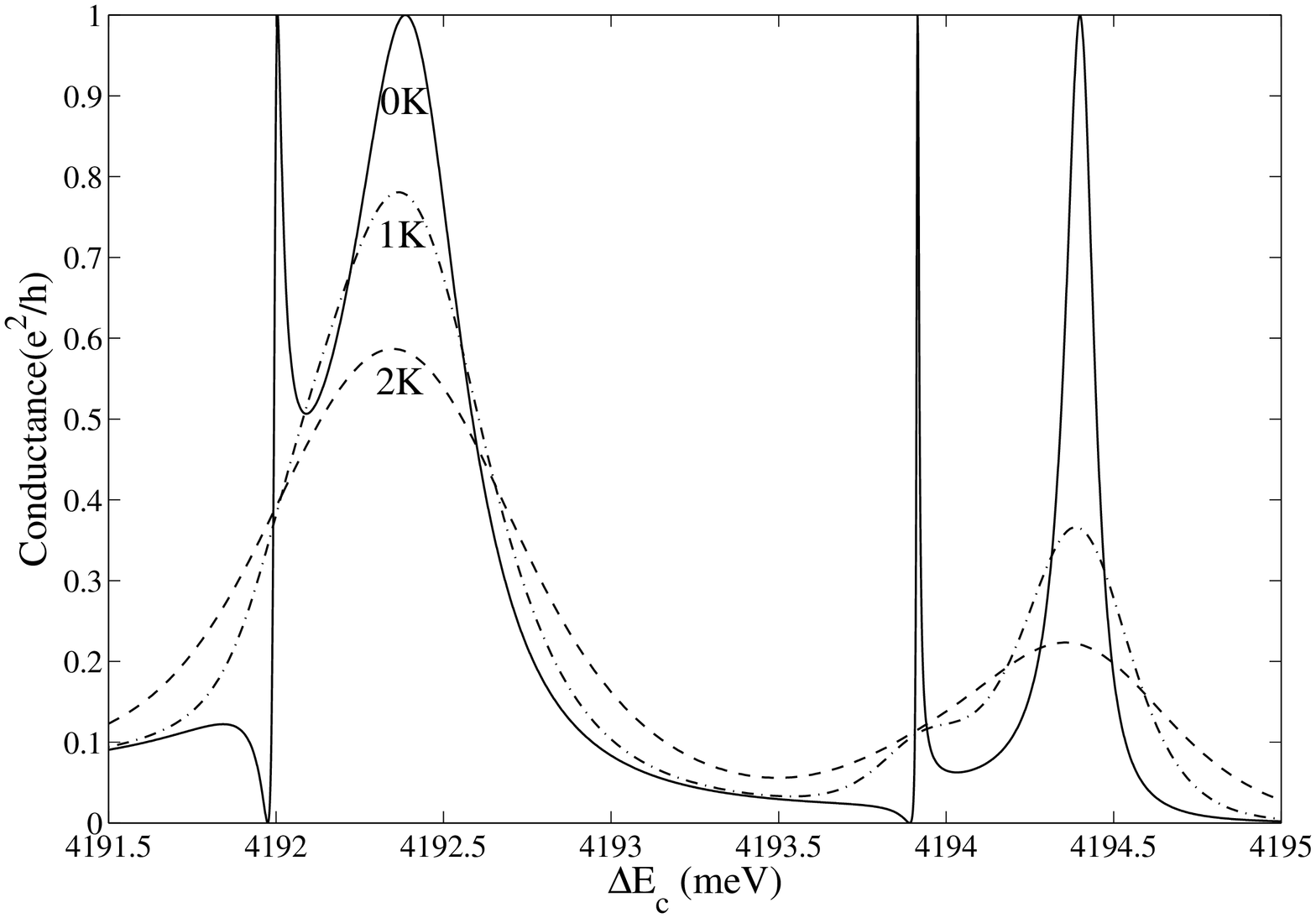,height=3.in,width=4.3in,angle=-90}
}
\end{figure}
\vskip 2in
\begin{center}
{\bf Figure 4(a)}
\end{center}

\newpage
\
\vskip 0.2in
\begin{figure}[h]
\centerline{\psfig{figure=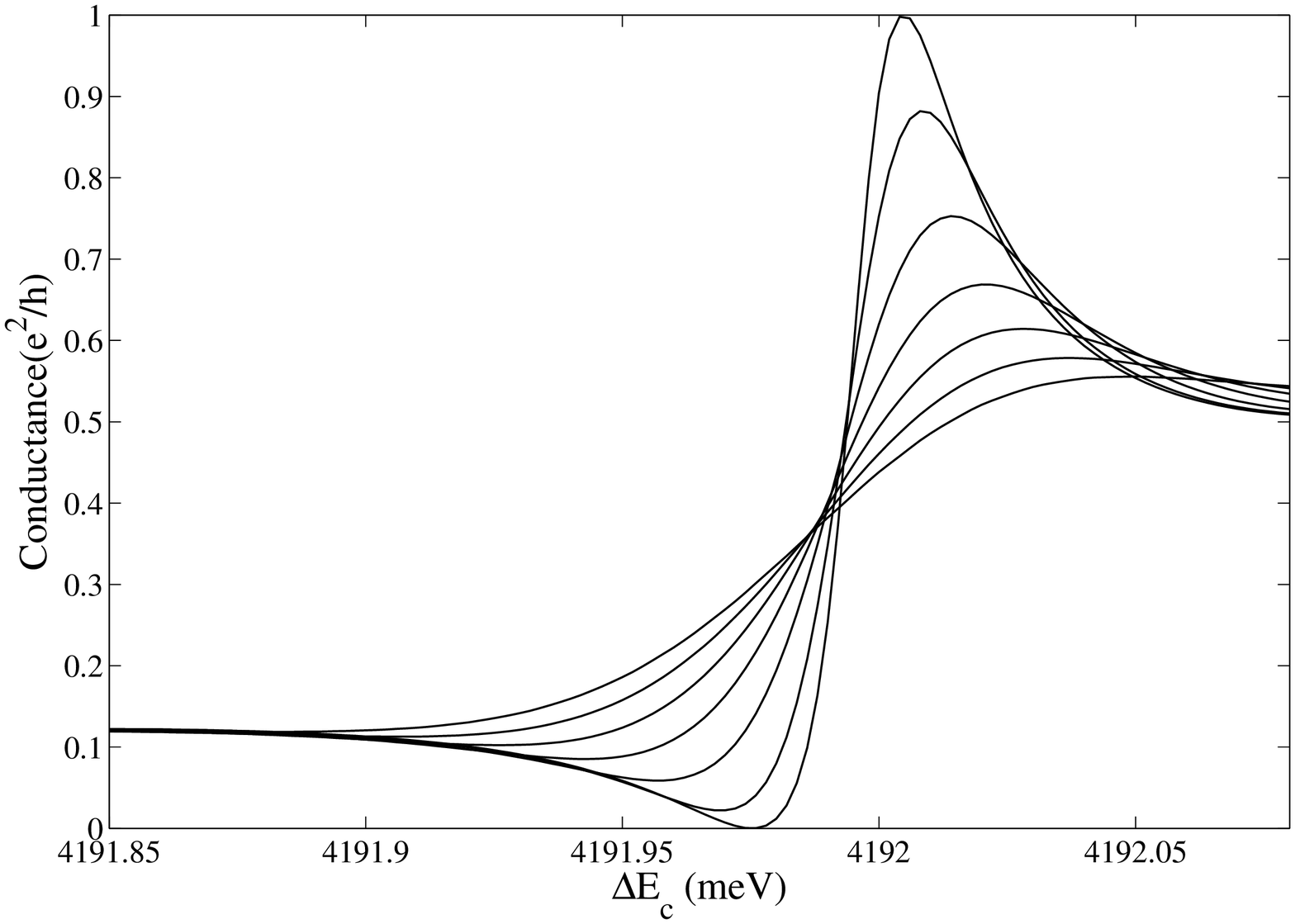,height=3.in,width=4.3in}}
\end{figure}
\vskip 2in
\begin{center}
{\bf Figure 4(b)}
\end{center}

\end{document}